\documentclass[english,aps,prb,twocolumn, superscriptaddress, floatfix, 10pt]{revtex4}
\usepackage[T1]{fontenc}
\usepackage[latin9]{inputenc}
\usepackage{amsmath}
\usepackage{graphicx}
\usepackage{amssymb}

\makeatletter

\@ifundefined{textcolor}{}
{
 \definecolor{BLACK}{gray}{0}
 \definecolor{WHITE}{gray}{1}
 \definecolor{RED}{rgb}{1,0,0}
 \definecolor{GREEN}{rgb}{0,1,0}
 \definecolor{BLUE}{rgb}{0,0,1}
 \definecolor{CYAN}{cmyk}{1,0,0,0}
 \definecolor{MAGENTA}{cmyk}{0,1,0,0}
 \definecolor{YELLOW}{cmyk}{0,0,1,0}
 }

\newcommand{\carbon}{$^{13}$C}
\newcommand{\siliconspin}{$^{29}$Si}

\newcommand{\tone}{$T_1$}

\makeatother

\usepackage{babel}

\begin{document}

\title{Decay of nuclear hyperpolarization in silicon microparticles}

\author{M. Lee}
\affiliation{Department of Physics, Harvard University, Cambridge, Massachusetts 02138, USA}
\author{M.~C. Cassidy}
\affiliation{School of Engineering and Applied Sciences, Harvard University, Cambridge, Massachusetts 02138, USA}
\author{C. Ramanathan}
\altaffiliation{Present address: Department of Physics and Astronomy, Dartmouth College, Hanover, New Hampshire 03755, USA}
\affiliation{Department of Nuclear Science and Engineering, Massachusetts Institute of Technology, Cambridge, Massachusetts 02139, USA}
\author{C.~M. Marcus}
 \altaffiliation{To whom correspondence should be addressed (marcus@harvard.edu)}
\affiliation{Department of Physics, Harvard University, Cambridge, Massachusetts 02138, USA}
\begin{abstract}
We investigate the low-field relaxation of nuclear hyperpolarization in undoped and highly doped silicon microparticles at room temperature following removal from high field. For nominally undoped particles, two relaxation time scales are identified for ambient fields above 0.2 mT. The slower, $T_{1,\rm s}$, is roughly independent of ambient field; the faster, $T_{1,\rm f}$, decreases with increasing ambient field. A model in which nuclear spin relaxation occurs at the particle surface via a two-electron mechanism is shown to be in good agreement with the experimental data, particularly the field-independence of  $T_{1,\rm s}$.  For boron-doped particles, a single relaxation time scale is observed. This suggests that for doped particles, mobile carriers and bulk ionized acceptor sites, rather than paramagnetic surface states, are the dominant relaxation mechanisms. Relaxation times for the undoped particles are not affected by tumbling in a liquid solution.
\end{abstract}
\maketitle

\section{Introduction}

Silicon has long been a staple of the microelectronics industry and so has been the subject of intense materials research. In recent years, nanoscale silicon has attracted attention due to its unique electronic and optical properties,\cite{Ledoux, Delley, Kim} and as a potential agent for medical imaging and drug delivery.\cite{Tasciotti, Park, Aptekar} Spin phenomena in silicon has been investigated using nuclear magnetic resonance (NMR) for over a half-century,\cite{Pines, Feher, McCamey, Yang, Bloembergen49, Bloembergen54, Shulman, Ladd05} while material properties and fabrication has continually developed. For instance, the nuclear spin-lattice relaxation time, \tone, provides information about dopants and impurities.\cite{Bloembergen49, Bloembergen54, Shulman} Notably, the low natural abundance (4.7\%) of spin-1/2 \siliconspin~nuclei in a lattice of zero spin nuclei leads to \tone~of many hours \cite{Shulman, Aptekar} and coherence times up to tens of seconds \cite{Ladd05} in undoped bulk crystalline silicon. 

These remarkable NMR properties have stimulated interest in silicon as a platform for solid-state quantum computing,\cite{Kane, Ladd02, Itoh05} and as a long-lived imaging agent for hyperpolarized magnetic resonance imaging (MRI).\cite{Aptekar} In particular, there have been renewed efforts to understand dynamic nuclear polarization (DNP), a process where saturation or pumping of paramagnetic impurity states by microwave fields or optical illumination can lead to nuclear spin polarizations orders of magnitude larger than thermal equilibrium polarizations.\cite{Lampel, Verhulst, Dementyev, Abragam, Hayashi09} Enhancing polarization by these methods is commonly referred to as hyperpolarization.

Application of silicon as a hyperpolarized imaging agent for MRI requires an understanding nuclear spin relaxation over a broad range of applied fields, from high fields where polarization is induced and imaged, to low fields where the agent is transferred from the polarizer and administered.   Previous detailed studies have focussed on low-temperatures, investigating the dependence of $T_1$ on applied magnetic field, doping\cite{Bagraev79b, Bagraev82a, Bagraev79a} and strain.\cite{Bagraev82b} It was found that $T_1$ increased with applied magnetic field, which was explained in terms of the field dependence of electron-nuclear dipole-dipole interaction. The nuclear $T_1$ was also shown to scale inversely with the concentration of mobile carriers.\cite{Bloembergen54, Shulman} In silicon nanoparticles, the observed increase in nuclear $T_1$ with increasing particle size was attributed to diffusion-mediated relaxation via defects at the particle surface.\cite{Aptekar}
Recent measurements have shown that the relaxation and coherence times of $^{31}$P donor-bound electron spin are dramatically reduced near the silicon surface due to P$_{\rm b}$ defects, in comparison to known values in the bulk.\cite{Paik} Studies of electrically\cite{McCamey} and optically\cite{Yang} detected hyperpolarization of $^{31}$P donor nuclear spins near the silicon surface, however, do not consider effects of the surface on nuclear spin relaxation.

In this paper, we investigate the decay of hyperpolarization of $^{29}$Si nuclei in silicon microparticles at room temperature and low ambient magnetic fields after removing the samples from a high-field polarizing environment. We find the decay is bi-exponential, with a slow time scale that is independent of ambient field, and a fast time scale that shows only a modest decrease with increasing ambient field. We develop a model of nuclear spin relaxation in silicon particles that takes into account the heterogeneous makeup of the sample, with direct nuclear relaxation occurring only near the surface, while nuclei in the particle core are relaxed indirectly by spin diffusion. This model extends a previous spin diffusion model, \cite{Dementyev} which predicts $T_1$ increasing with the square of the particle diameter. We find that the weak magnetic field dependence observed experimentally is inconsistent with a simple extended spin-diffusion model based on relaxation on individual bound electrons. However, by generalizing the model further to include nuclear spin relaxation mediated by pairs of dipolar-coupled electrons, the essentially field-independent relaxation is recovered by the model. \cite{Terblanche, Khut, Duijvestijn, Ramanathan} The experimental methods presented here, as well as the use of magnetic field dependence of nuclear $T_1$ as a probe for testing the relative importance of these different model processes, have precedent in a similar investigation performed on diamond and its $^{13}$C nuclear spin relaxation.\cite{Terblanche,Shabanova,Panich}

\section{Samples and Measurements}

Microparticles made from nominally undoped and boron-doped  ball-milled Si wafers were investigated. As shown previously \cite{Aptekar}, undoped Si microparticles have very long high-field relaxation times ($T_1 \sim10^4$ s), relevant for hyperpolarized MRI. Undoped samples were produced by ball-milling high-resistivity (> 10\,k$\Omega\cdot$cm) float-zone Si wafers (Silicon Quest International) followed by size separation by centrifugal sedimentation, yielding a mean particle diameter of 5 $\mu$m.\cite{Aptekar} Boron-doped Si microparticles with doping density $\sim 5\times10^{18}\mbox{ cm}^{-3}$ have much faster nuclear spin relaxation, $T_1$~$\sim10^2$ s at high magnetic field. The boron-doped microparticles were produced by ball-milling  Czochralski-grown Si wafers (Virginia Semiconductor, resistivity 0.01-0.02~$\Omega\cdot$cm) followed by size separation, yielding a mean particle diameter on the order of 10~$\mu$m.  Except where noted, samples consisted of a 0.1~mL teflon tube filled with particles. All NMR measurements were carried out at $B_0 = 2.9$\,T using a custom-built probe and spectrometer. 

Polarization at 2.9\,T following a saturation sequence of sixteen $\pi/2$ pulses was measured as a function of time, $t_{\rm pol}$, using a Carr-Purcell-Meiboom-Gill (CPMG) sequence \cite{Carr, Meiboom, Li}, $\left(\pi/2\right)_{X}-\left[t/2-\left(\pi\right)_{Y}-t/2-\mbox{echo}\right]^{n}$ with $t=1\mbox{ ms}$ and $n=400$. After each data point, the sample was re-saturated and the measurement repeated. As seen in Figs.~1(a) and 4(a), the build-up of polarization is well described by a single exponential function of $t_{\rm pol}$ for both undoped and doped samples.

Depolarization at low ambient fields was measured after first polarizing at 2.9\,T for 8 h ($\sim 3\,T_1$ at 2.9\,T), then raising the probe out of the magnet bore to a position where ambient (or holding) fields of 0.2~mT, 6~mT, 130~mT, and 300~mT had been previously calibrated using a Lakeshore 460 gaussmeter.  In addition, a nominal zero-field measurement used a commercial zero-gauss chamber (Lakeshore 4060) with ambient field below $1~\mu$T. Following depolarization in ambient field, the sample was returned to field center (2.9\,T) and the remaining nuclear polarization was measured using the CPMG sequence described above. The transit time between positions was $\sim~\!\!\!5$~s, fast compared to $T_1$ but slow compared to nuclear Larmor times. The sample was fully repolarized following each data point.

In addition to NMR measurements, room temperature electron spin resonance (ESR) measurements were carried out to characterize electronic defects and carriers of the two sample types. The ESR spectrum of the undoped sample showed a single peak at $B = 324$ mT measured at ESR frequency $f = 9.099$ GHz, corresponding to a g-factor centered at 2.006, with a linewidth of 0.47~mT. The ESR spectrum of the boron-doped sample showed a single peak at $B = 336$ mT measured at ESR frequency $f = 9.444$ GHz, again corresponding to a g-factor of 2.006.
A g-factor of 2.006 is consistent with reported values for P$_{\rm b}$ defects at the Si-SiO$_2$ interface, and is not consistent with g-factors of other common dopants or oxide states.\cite{Nishi, Caplan}

\begin{figure}
\includegraphics[width=8.6cm]{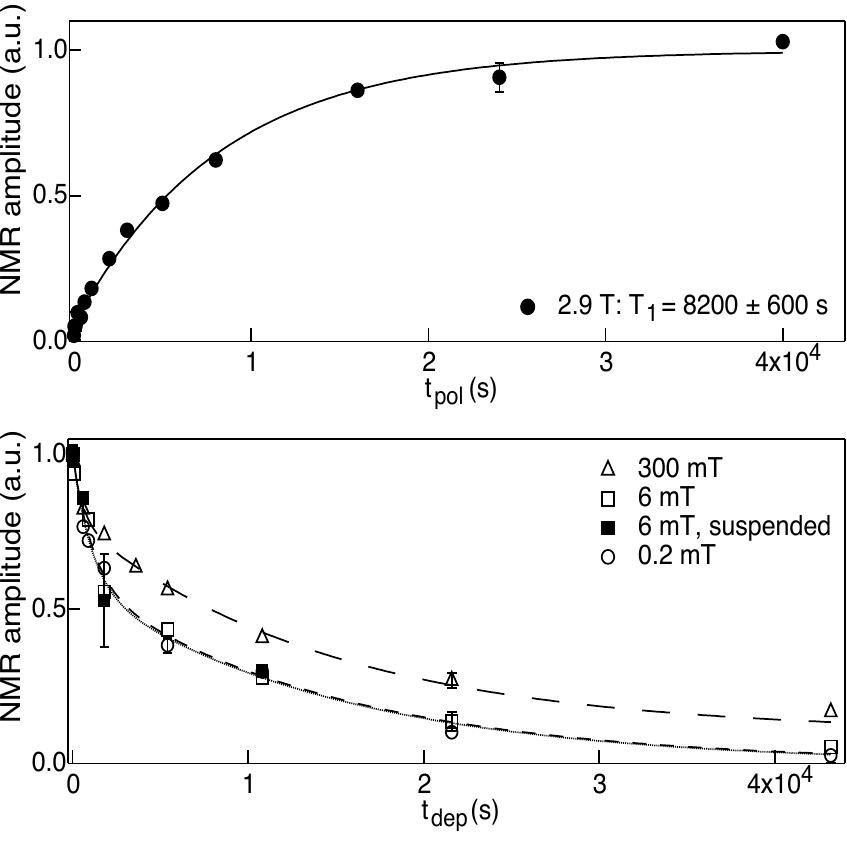}
\caption{Relaxation of nuclear spin polarization of undoped silicon particles. (a) Recovery of spin polarization following saturation at $B_0=$ 2.9~T. 
(b) Decay of nuclear magnetization at $B_{\rm dep}=$ 0.2~mT (circles), 6~mT (open squares), and 300~mT (triangles), after polarization at 2.9~T, and same data for particles suspended in methanol at 6~mT (filled squares). Curves show best fits to Eq.~1. No signal was observed after 1 second for data taken in the zero-field chamber (< 1$\mu$T).}
\end{figure}

\subsection{ Undoped Microparticles: Measurements}

Polarization and depolarization data for the undoped microparticles are shown in Fig.~1. Polarization is well described by a simple exponential,  $P=P_{0,B_0}\left(1-e^{-t_{\rm pol}/T_{1}}\right)$, with best-fit value $T_{1}=8200\pm600\mbox{ s}$, consistent with previous measurements.\cite{Aptekar} Equilibrium polarization is small, $P_{0,B_0}=\tanh[(\hbar \gamma_{n}B_{0})/(2k_B T)]= 2\times 10^{-6}$. Here, $k_B$ is Boltzmann constant, $T=300\mbox{ K}$ is room temperature, and $\gamma_{n}=5.31\times 10^7\,s^{-1}\,{\rm T}^{-1}$ is the nuclear gyromagnetic ratio for $^{29}$Si.

In contrast to the build-up of polarization at 2.9 T, depolarization at low ambient fields, $B_{\rm dep}$,  decays with two distinct time scales, which we characterize using a bi-exponential form,
\begin{align} P & =P_{0,B_{\rm dep}}+\left(P_{0,B_{0}}-P_{0,B_{\rm dep}}\right) \\
 & \qquad\times\left((1-\alpha)e^{-t_{\rm dep}/T_{1,\rm f}}+\alpha e^{-t_{\rm dep}/T_{1,\rm s}}\right), \notag
\label{eq:biexp} \end{align}
where $T_{1,\rm f}$ and $T_{1,\rm s}$ are fast and slow relaxation times, and $\alpha$ is the fraction of spins whose polarization decays slowly. Best-fit values of $\alpha$, $T_{1,\rm f}$ and $T_{1,\rm s}$ are shown in Table \ref{table:nonlin}. Note that the slow relaxation time depends very weakly on $B_{\rm dep}$, with $T_{1,\rm s} \sim 1.4\,$-$\,1.5\times10^3$~s in all cases. In contrast, fast relaxation becomes somewhat faster (shorter $T_{1,\rm f}$) with increasing $B_{\rm dep}$, while the fraction of slow relaxers increases. 

Inside the zero-gauss chamber the residual ambient magnetic field is specified to be below the few-$\mu {\rm T}$ scale, much weaker than the typical nuclear dipolar field, $B_{\rm n, dd} \sim\mu_{0}\mu_{n}/(4\pi a^{3})=0.08\mbox{ mT}$, where $a\sim4$~\AA~is the mean separation between randomly distributed $^{29}$Si nuclei. At such extremely low fields, spin transitions will occur that are forbidden when $B > B_n$ by the conservation of nuclear Zeeman energy, and nuclear spin relaxation is as fast as decoherence.\cite{AbragamBook} As expected, we find that in the zero-gauss chamber nuclear polarization decays very quickly, with no visible signal for $t_{\rm dep} >1\mbox{ s}$.

To investigate the effect of microparticle tumbling on depolarization, a fluid mixture of one part Si microparticles and four parts methanol by weight was compared to a sample of packed dry powder. Depolarization times at 6~mT ambient field showed no significant difference in relaxation times.

\begin{table}
\centering
\begin{tabular}{r l c c c}
\hline\hline
$B_{\rm dep}$ & &$\alpha$ & $T_{1,\rm f}$ (s) & $T_{1,\rm s}$ (s) \\ 
\hline
300 mT & &$0.765\pm0.027$ & $531\pm179$ & $13850\pm1030$\\
6 mT & &$0.593\pm0.050$ & $903\pm232$ & $15310\pm2070$\\
0.2 mT & &$0.591\pm0.120$ & $1124\pm544$ & $14910\pm5360$\\
< 1 $\mu$T & &\multicolumn{3}{c}{$<1$}\\
\hline\hline
\end{tabular}
\caption{Exponential weights and relaxation times fitting to experimental data. $T_{1,\rm f}$ and $T_{1,\rm s}$ are fast and slow relaxation times with which two additive components of the spin polarization decay. $\alpha$ is the relative amplitude of the component which decays at $T_{1,\rm s}$.}
\label{table:nonlin}
\end{table}

\begin{figure}[b]
\centering{}\includegraphics[width=8.6cm]{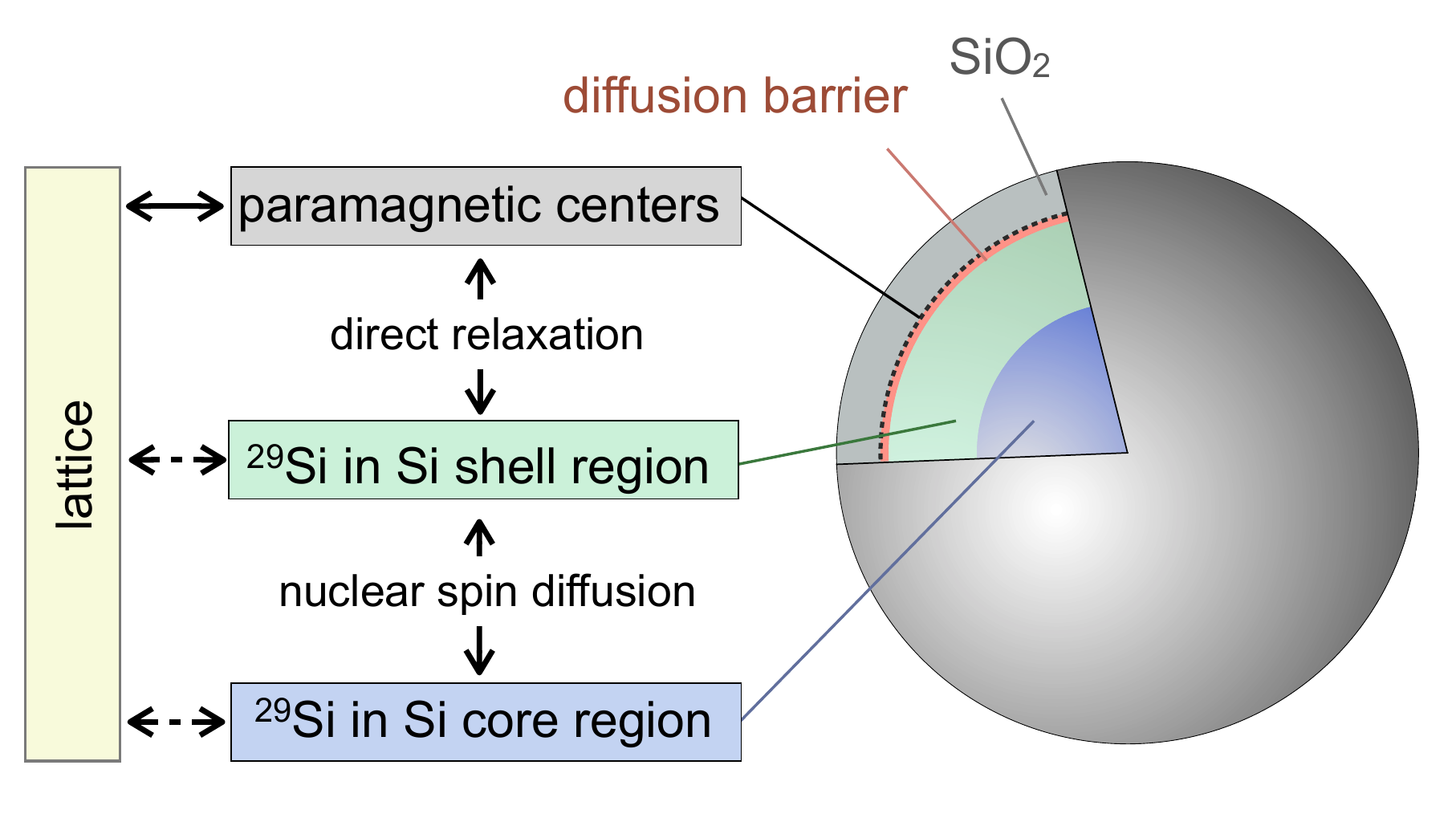}
\caption{Model showing the spin reservoirs and relaxation pathways in undoped silicon particles. Nuclei in the core region of the particle relax by transferring their magnetization to the nuclei in the shell region, nearer to the surface, by spin diffusion. The nuclei near the surface can relax quickly due to strong dipolar coupling to paramagnetic defects at the Si/SiO$_2$ interface.}
\label{Model1a}
\end{figure}

\subsection{Undoped Microparticles: Model}

We interpret NMR and ESR data for the undoped microparticles within a model shown schematically in Fig.~2, comprising nuclear dipolar diffusion in a core region and relaxation via paramagnetic sites in a shell region near the Si/SiO$_2$ interface. In undoped silicon, direct coupling of \siliconspin~nuclear spins to phonons is weak \cite{Bloembergen49, Khut}, and dipolar coupling to paramagnetic impurities and defects typically dominates nuclear spin relaxation. Within our model, the bi-exponential form of Eq.~1 reflects the existence of two populations of nuclear spins within each microparticle: a fraction $\alpha$, located in the core of the particle, has a long spin relaxation time $T_{1,\rm s}$ mediated by nuclear spin diffusion; the remaining fraction $(1-\alpha)$, located within a shell near the Si/SiO$_2$ interface, has a short relaxation time $T_{1,\rm f}$ mediated by electron-nuclear spin interaction associated with paramagnetic centers at the Si-SiO$_2$ interface. 

For $B_{\rm dep} > B_{\rm n,dd}$, the nuclear dipolar spin diffusion rate is independent of magnetic field,\cite{Abragam} and is well described by a diffusion constant $D=Wa^{2} \sim a^{2}/(50\,T_2)$, where $W$ is the probability of a flip-flop transition between nuclei due to dipole-dipole interaction, and $T_2$ is the nuclear decoherence time.\cite{Khut}  This is consistent with our observation that $T_{1,\rm s}$ is roughly independent of $B_{\rm dep}$ across a broad range of values.

The thickness of the shell region is set by the nuclear spin diffusion radius, $\beta=(C/D)^{1/4}$, where $C$ is a constant describing the nature of the dipolar interaction occurring between the nuclei and electrons.
Nuclei situated at a distance $r$ from the electrons at the particle surface may relax through a dipolar interaction with the electrons at a rate $\propto C/r^6$.\cite{Khut}

We consider two physical mechanisms of relaxation in shell region, one in which each nuclear spin interacts with a single electron spin, and the other in which each nuclear spin interacts with a pair of electron spins. The two-electron model captures a crucial physical mechanism by allowing energy matching between electron-spin-pair flip-flops and nuclear spins flips in a magnetic field, leading to efficient nuclear relaxation with weak dependence on magnetic field.   

Modeling nuclear spin relaxation as arising from interaction with individual paramagetic centers is appropriate for sparsely distributed impurities. This regime has been studied experimentally, for instance, in diamagnetic crystals such as LiF \cite{Cox, Goldman} and $\rm CaF_{2}.$\cite{Blumberg} In this case, the orientation-averaged dipole interaction strength is given by \begin{align} C_{(1e)} =\frac{3}{10}\frac{\hbar^{2}\gamma_{e}^{2}}{T_{1e}B_{\rm dep}^{2}}, \end{align} where $\gamma_{e}$ is the gyromagnetic ratio of the electron and $T_{1e}$ the electron spin-lattice relaxation time.\cite{Khut} This model yields a magnetic field dependence of diffusion-limited relaxation ($R \gg \beta$) at least as strong as $T_{1, \rm s} \propto B_{\rm dep}^{1/2}$ because the diffusion radius $\beta$ and relaxation rate $\propto\,C$ depend explicitly on field, even when $D$ is field independent.\cite{Khut, Blumberg, Lowe} This prediction is incompatible with our experimental results, which show a much weaker field dependence. 

In light of this inconsistency, we consider a model of nuclear relaxation at the particle surface that includes three-spin processes involving pairs of interacting electrons, which gives \begin{align} C_{(2e)} = & \frac{3}{10}\frac{\hbar^{2}\gamma_{e}^{2}}{B_{\rm dep}^{2}}  \frac{B_{\rm dep}^{2}\gamma_{n}^{2}T_{2e}}{1+ B_{\rm dep}^{2}\gamma_{n}^{2}T_{2e}^2} \int_{-\infty}^{\infty} \! \frac{g(\omega)g(\omega-\omega_n)}{g(0)} \, \mathrm{d}\omega, \end{align} where $T_{2e}$ is the electron spin-spin coupling and $g(\omega)$ is the normalized electron absorption lineshape function.\cite {Duijvestijn, Terblanche} This model has been previously applied to systems with more concentrated paramagnetic impurities, including  $\mbox{La}_{2}\mbox{Mg}_{3}(\mbox{NO}_{3})_{12} \cdot 24\mbox{H}_{2}\mbox{O}$ \cite{Heuvel, Wenckeback} and more recently {$^{13}$C} nuclear spin relaxation in diamond.\cite{Terblanche, Shabanova} This model accounts for flip-flop transitions between nearby electron pairs, occurring on a time scale $T_{2e} \ll T_{1e}$, which provide the fluctuating magnetic field that can flip nuclear spins. That is, when the dipolar coupling of electron pairs matches the nuclear Zeeman energy, a three-spin interaction can occur that exchanges the spins of the two electrons while flipping a nuclear spin. \cite{Wollan}  This process depends on the density of transitions between electronic dipolar energy states that match the nuclear Zeeman energy. For a Lorentzian electron lineshape, the integral in (3), which describes the probability of finding two electrons within the ESR line differing in frequency by $\omega_{n}$, can be replaced by $2/(4+\omega_{n}^{2}T_{2e}^{2})$. For low $B_{\rm dep}$, such that $\gamma_n B_{\rm dep} < \gamma_e B_{\rm e, dd} \sim T_{2,\rm e}^{-1}$, where $B_{\rm e,dd}$ is the electronic dipole field felt by a typical nucleus, the density of nuclear-spin-flip transitions will be independent of $B_{\rm dep}$, hence the rate of direct nuclear spin relaxation by this process will also be independent of $B_{\rm dep}$.\cite{Duijvestijn, Terblanche}

We infer an upper bound for $B_{\rm e, dd}$ from the measured ESR linewidth of 0.47~mT, which gives   $\gamma_n^{-1} \gamma_e B_{\rm e,dd} \sim 1.7$~T for the scale of $B_{\rm dep}$ below which nuclear $T_1$ should be roughly field independent within this three-spin model. The low end of the field-independent range is set by the nuclear dipolar field $B_{\rm n, dd} \sim 0.08$~mT. This range is consistent with the experimental observation that $T_{1,\rm s}$ is roughly field independent from 0.2~mT to 300~mT. Weak field dependences of $T_{1,\rm f}$ and $\alpha$ are likely due to the weak field dependence of $T_{1e}$.

Very close to each paramagnetic impurity, nuclear spin diffusion is suppressed, creating a barrier to diffusion of radius $b = a(\hbar\gamma_{e}^2 B_{\rm dep} / \gamma_{n}2k_{\rm B}T)^{1/4}$, due to gradients in the nuclear Larmor frequency. \cite{Khut} This effect is included in the simulations (described below), but because $\beta \gg b$ it only weakly affects overall relaxation times.

\begin{figure}
\centering{}\includegraphics[width=8.6cm]{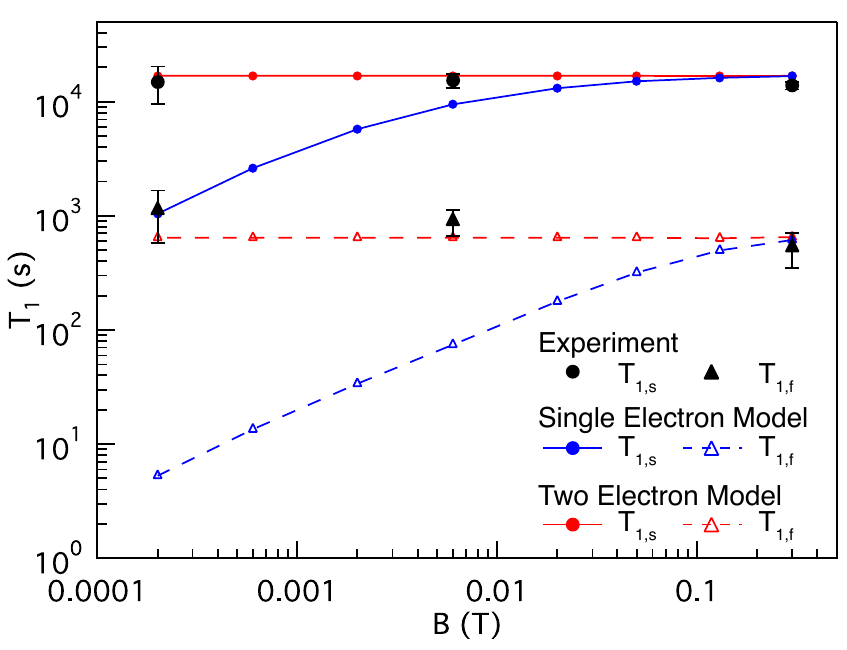}
\caption{Experimental slow (black circle)  and fast (black triangle) relaxation times, $T_{1,\rm s}$ and $T_{1,\rm f}$, from Table~\ref{table:nonlin}, for bi-exponential nuclear spin relaxation, along with simulation results for a model that includes either one-electron or two-electron processes.}
\label{Models}
\end{figure}

We have simulated nuclear spin relaxation as a function of ambient field for a spherical silicon microparticle with native $^{29}$Si concentration and paramagnetic sites at the surface. For both one and two electron processes, the average polarization $P(t_{\rm dep})$ is approximately bi-exponential, similar to the experimental data. Fits to the relaxation curves gave $T_{1,\rm s}$ and $T_{1, \rm f}$ values shown in Fig.~\ref{Models}. The particle diameter was taken to be 700~nm, smaller than the actual mean particle size. This compensated the spherical assumption, which gave too small of a surface where defects reside. Other parameters in the model were nuclear $T_2$ of 10~ms (and corresponding nuclear spin diffusion constant 0.3~nm$^2/$s), electron $T_{1e}$ of 200~ns, and electron $T_{2e}$ of 25~ns. Figure \ref{Models} shows the striking contrast in field dependence between the two models, and that the three-spin process compares relatively well with experiment.

\subsection{Results for Boron-doped Microparticles}

Nuclear polarization and depolarization of boron-doped Si microparticles are shown in Fig.~4. Both $P(t_{\rm pol})$ and $P(t_{\rm dep})$ are well fit by single-exponential relaxation curves with time constants $T_1$ that increases with increasing ambient field. The reduction of $T_1$ by 1-2 orders of magnitude in the doped case reflects the dominant contribution of nuclear relaxation from mobile carriers,\cite{Bloembergen54} as well as from ionized acceptors distributed uniformly through the particle instead of just at the surface (Fig.~5), bypassing the relatively slow nuclear spin diffusion process. For the specified dopant density, the average distance between dopants is $\sim2.5$~nm and the Bohr radius $\sim1.3$~nm\cite{Tanaka}, both an order of magnitude less than the diffusion radius for both single electron and two-electron processes across the range of ambient fields under study. The contribution of mobile carriers can be expressed as an additional relaxation rate which is independent of magnetic field, $T_{1, \rm carriers}^{-1} \propto N_0 A^2$ where $N_{0}$ is concentration of mobile carriers, and $A$ is the strength of the hyperfine coupling. \cite{Bloembergen54} The contribution to nuclear relaxation by ionized acceptors as well as surface paramagnetic defects at the surface are responsible for the mild magnetic field dependence of $T_1$ seen in Fig.~4(b). We were unable to observe an ESR signal from the mobile carriers,  presumably due to their short relaxation times leading to broad broad ESR lines. 

\begin{figure}[t]
\includegraphics[width=8.6cm]{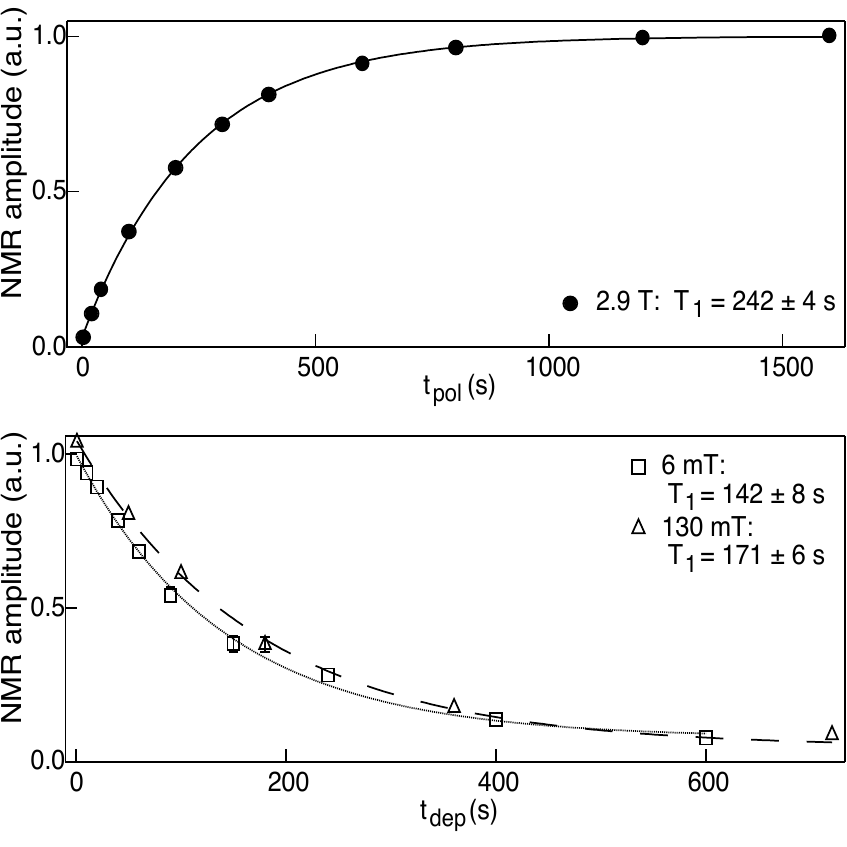}
\caption{Nuclear spin lattice relaxation in boron-doped silicon microparticles.
(a) Recovery of spin polarization after saturation at $B_0=$ 2.9~T.
(b) Decay of nuclear magnetization at $B_{\rm dep}=$ 6~mT (open square), and 130~mT (triangle), after
polarization at 2.9~T. Error bars in (a) are displayed but smaller than the data points shown.}
\label{Figure3}
\end{figure}

\begin{figure}
\centering{}\includegraphics[width=8.6cm]{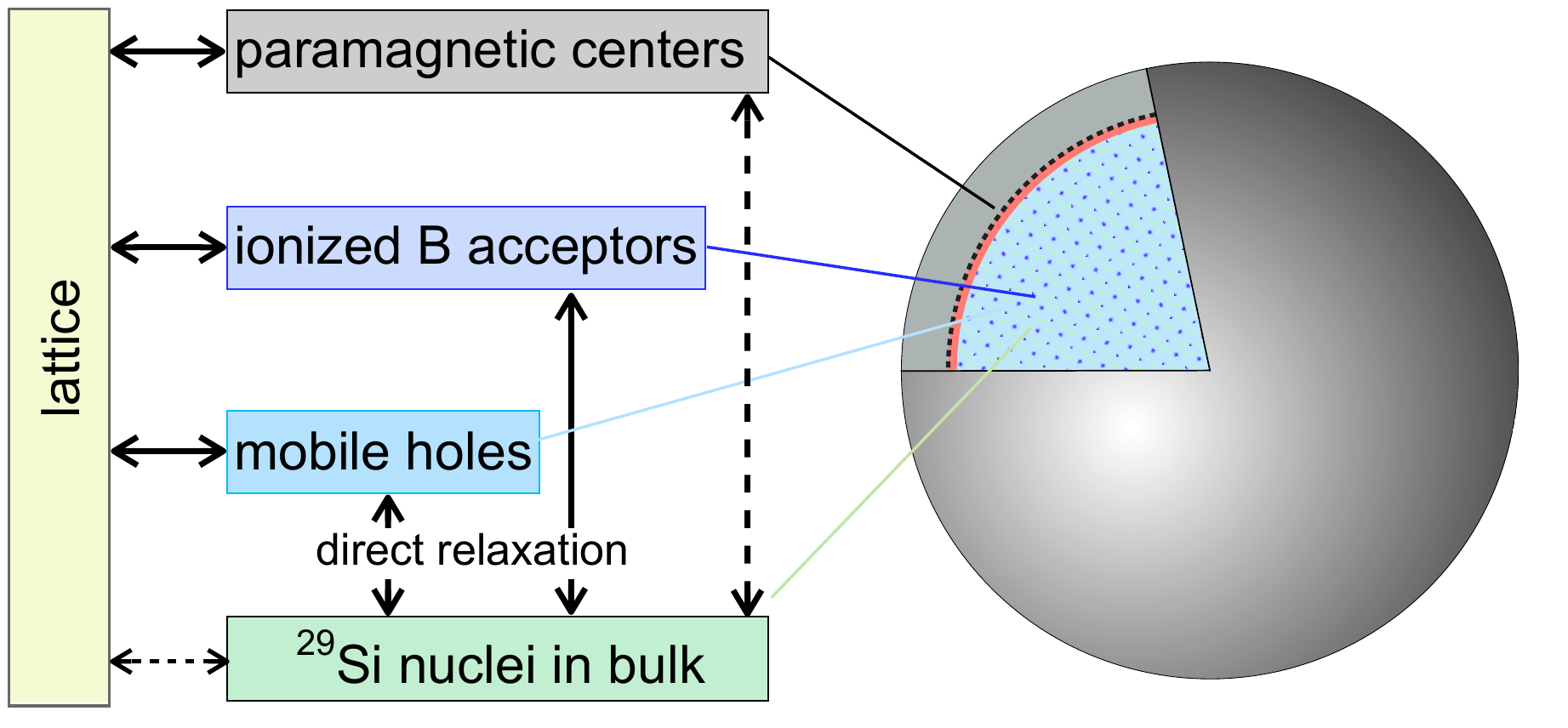}
\caption{Rate model showing dominant nuclear relaxation pathways of direct relaxation on mobile holes and short diffusion to ionized boron acceptors throughout the bulk of the particle. Long distance spin diffusion to surface impurities is relatively slow and direct coupling of the nuclei to the lattice is weak.}
\label{Model1b}
\end{figure}

\section{Conclusion}

We have investigated the relaxation of polarization of natural-abundance \siliconspin~nuclei in undoped and doped silicon microparticle samples as a function of ambient magnetic fields at room temperature.  In undoped micron-scale particles, the decay of nuclear polarization from a hyperpolarized value was measured from microtesla to 0.3 T. For fields stronger than the nuclear dipolar field, a bi-exponential decay of nuclear magnetization was observed, with a fast component of $\sim 10^2$ s that depends weakly on field and a slow component of $\sim 10^4$ s that is roughly field independent. The relative amplitude of the slowly decaying component increases with the magnetic field, but only slightly. Bi-exponential relaxation suggests the presence of two nuclear spin baths distinguished by their proximity to paramagnetic impurities at the particle surface. The timescales of nuclear spin relaxation are largely independent of ambient field $B_{\rm dep}$ from 0.2 to 300~mT. This independence from $B_{\rm dep}$ is quantitatively consistent with a model of nuclear spin relaxation dominated by a three-spin mechanism in which flip-flop transitions of two electrons at the nuclear Larmor frequency flip a nuclear spin.  In highly doped silicon, simple exponential relaxation with a faster $T_1$ of order a few hundred seconds is observed.

These results can be compared with similar experiments performed in diamond, another material with dilute (1.1~\%) spin-1/2 nuclei among a majority of zero-spin nuclei in an fcc lattice. \cite{Hoch, Reynhardt, Terblanche}. Measurement of the nuclear $T_1$ in natural diamond as a function of magnetic field indicate a contribution of three-spin processes to \carbon~spin relaxation.\cite{Terblanche} Other investigations of nuclear spin relaxation in \carbon-enriched diamond\cite{Shabanova} and in nanocrystalline diamond\cite{Panich} have shown that the relaxation of nuclear magnetization was not well described by a simple exponential approach to equilibrium. The NMR studies in nanocrystalline diamond revealed that the particles consist of a crystalline core and a surrounding shell. All these observations in diamond are similar to our results in silicon microparticles.

Finally, we comment that multi-hour and field-independent nuclear relaxation times for a substantial fraction of the nuclear spin population, including for particles in a fluid suspension, is important for for the use  of silicon microparticles as a hyperpolarized MRI imaging agent. \\

\section*{Acknowledgments}

We thank David Cory for generous access to the ESR spectrometer. We acknowledge support from the National Science Foundation under NSF-0702295, the BISH Program (CBET-0933015),and the Harvard NSF Nanoscale Science and Engineering Center. Fabrication used the Harvard Center for Nanoscale Systems (CNS), an NSF National Nanotechnology Infrastructure Network (NNIN) site (ECS 0335765). ML acknowledges support from Samsung.

\end{document}